\documentclass[fleqn,usenatbib]{mnras}

\usepackage{newtxtext,newtxmath}

\usepackage[T1]{fontenc}

\DeclareRobustCommand{\VAN}[3]{#2}
\let\VANthebibliography\thebibliography
\def\thebibliography{\DeclareRobustCommand{\VAN}[3]{##3}\VANthebibliography}

\usepackage{graphicx}	
\usepackage{amsmath}	
\usepackage{fix-cm}
\usepackage{hyperref}

\title[Very Low Mass Eclipsing Binary System CM Draconis]{Comprehensive analysis of CM Draconis: eclipse timing variations driven by either a third body or stellar magnetic activity}

\author[B. Kalomeni and K. Yakut]{
Belinda Kalomeni$^{1}$\thanks{E-mail:belinda.kalomeni@ege.edu.tr (BK)}
and
Kadri Yakut$^{1,2}$
\\
$^{1}$Department of Astronomy and Space Sciences, Faculty of Science, Ege University, 35100, {\.I}zmir, Türkiye\\
$^{2}$Institute of Astronomy, The Observatories, Madingley Road, Cambridge CB3 OHA, UK\\
}

\date{Accepted XXX. Received YYY; in original form ZZZ}

\pubyear{2025}

\begin{document}
\label{firstpage}
\pagerange{\pageref{firstpage}--\pageref{lastpage}}
\maketitle

\begin{abstract}
The CM Draconis system is a well-studied, double-lined spectroscopic binary that is totally eclipsing and exhibits strong magnetic activity. Nearly one million photometric measurements have been collected across multiple wavelengths over more than half a century. In addition to showing frequent flare activity and apsidal motion, CM Dra also hosts a distant white dwarf and has been proposed to hharbour a Jupiter-sized circumbinary companion. At only 47 light-years from Earth, it remains one of the most observationally rich and dynamically intriguing low-mass binary systems.
We present a comprehensive photometric and spectroscopic analysis of the system using new ground-based observations and data from 19 sectors of the \textit{TESS} mission. We derive precise fundamental parameters for both components: $M_1 = 0.2307 \pm 0.0008\,M_\odot$, $M_2 = 0.2136 \pm 0.0008\,M_\odot$, $R_1 = 0.2638 \pm 0.0011\,R_\odot$, $R_2 = 0.2458 \pm 0.0010\,R_\odot$, $L_1 = 0.0060 \pm 0.0005\,L_\odot$, and $L_2 = 0.0050 \pm 0.0004\,L_\odot$. The derived distance ($14.4 \pm 0.6$ pc) is consistent with \textit{Gaia} DR3 measurements. Eclipse timing variations (ETVs) spanning over five decades were analysed in detail. A long-period ($\sim$56 yr) modulation was identified, which may be attributed either to the light-time effect of a possible circumbinary companion or to magnetic activity cycles. While the Bayesian Information Criterion statistically favors the model involving a light-time effect from a planetary companion, stellar activity remains a viable alternative that cannot yet be ruled out. Our results demonstrate that CM Dra is a valuable test case for studying both stellar activity and the potential presence of circumbinary companions in multiple-star systems. Continued long-term monitoring will be essential to distinguish between these competing scenarios.
\end{abstract}




\section{Introduction}
\label{sec:intro}

There are numerous extrasolar planet research projects from ground-based to space-based telescopes.
TESS and Kepler projects are some of the foremost projects; they have obtained many fruitful results in this area.
Kepler and TESS have detected more than 12000 exoplanet candidates, a quarter of which have been confirmed, with orbital periods of less than a month in most cases \citep{Borucki2010Sci...327..977B,2014Natur.513..336L,TESS2015JATIS...1a4003R}.
Transit and radial velocity variation methods are used in detecting planets orbiting a single star like our Sun.
If a planet orbits a binary star, orbiting two Suns, the weak planetary perturbation perturbs the orbit of the binary system. This perturbation provides an estimation of the orbital elements of the planetary component by combining the observations extended over the years, as presented in this study.

CM Dra is a close binary system consisting of cool, active, late-type (\texttt{dM4.5+dM4.5}) \textit{almost} twin main-sequence stars.
The system is a detached double-lined eclipsing binary system and has been classified as a BY Dra-type variable by \cite{1988AJ.....95..887C}.
Both photometric and spectroscopic observations of CM Dra revealed two low-mass, late-type stars with an orbital period of 1.27 days \citep{1977ApJ...218..444L}.
The space velocity of CM Dra suggests Pop~II components by \cite{1977ApJ...218..444L}. 
Although early analyses based on spectral energy distribution suggested a metal-poor nature \citep{1997MNRAS.291..780V,2002MNRAS.329..290V}, more recent spectroscopic studies consistently report a mildly subsolar metallicity in the range $\textrm{[Fe/H]} = -0.30 \pm 0.12$ \citep{2012ApJ...760L...9T}. This slightly metal-poor composition is crucial for modelling the structure and evolution of CM Dra’s low-mass components \citep{2009ApJ...691.1400M}.

There are many evolutionary models of CM Dra in the literature. 
As in the general case of M dwarfs, there is a discrepancy between the observed radius and the radius suggested by the models \citep{1995ApJ...451L..29C,1996ApJ...456..356M,2002ApJ...567.1140T,2009ApJ...691.1400M,2012MNRAS.422.2255S,2012MNRAS.421.3084M}. Recent studies, such as \citep{Martin2024}, have provided improved measurements of the fundamental parameters of CM Dra, accounting for radius inflation due to magnetic activity. Studies in the literature suggest that the age of the system is between 3 and 5 Gyr. The discrepancy between the radius suggested by the model and the observed radius is about 6\% in the case of CM Dra.
In a recent study of the Galactic space velocity of CM Dra and the white dwarf WD 1633+572, they fall in the same region \citep{2014A&A...571A..70F},
The age of CM Dra and WD 1633+572 is estimated to be $8.5\mp3.5$ Gyr, higher than previous studies.
This age of the system suggests that the difference between the radius proposed by the model and the observed radius is 2\%.

The system has been observed various times both with photometric and spectroscopic observations.
The absolute orbital elements of CM Dra were given by some studies \citep{1977ApJ...218..444L,2004ARep...48..751K,2009ApJ...691.1400M,2015IAUGA..2258029K}.
The system exhibits a small but non-zero eccentricity. Although such close binary systems are expected to have circularized orbits due to tidal interactions \citep{2008EAS....29....1M}, this residual eccentricity may indicate the gravitational influence of a third body orbiting CM Dra. In addition, flaring activity in CM Dra has been reported in earlier studies \citep{2007IBVS.5789....1N}.
The brightness variations observed between orbital phases 0.28 and 0.60, corresponding to the maximum light of CM Dra during 1977–1997, were explained by a large, long-lived asymmetric polar spot \citep{2004ARep...48..751K}. \cite{Martin2024} measured 125 flares that the system shows well and suggested that these flares may be preferentially polar. As they mentioned, this could have positive implications for the habitability of planets orbiting systems like CM Dra. This work will be discussed in more detail in Section \ref{sec:activity}.

The potential of eclipsing binary systems for exoplanet detection has been recognized in theoretical studies dating back several decades \citep[see][for details]{Schneider1991,Schneider1994}. In the early stages of research, it was demonstrated that the precession of a planet's orbit, due to the flattening of the central binary, could significantly enhance transit probabilities, even for planets on inclined orbits. Furthermore, \cite{Borucki1984} emphasized the advantages of photometric transit detection, noting that large planets produce measurable flux reductions and distinct colour variations, which could serve as a verification mechanism for planetary transits. These foundational studies laid the groundwork for modern space-based transit surveys and highlighted the viability of eclipsing binaries as prime targets for circumbinary planet (CBP) searches.

While the long-term orbital period change method offers a promising approach for detecting additional companions in binary systems, it inherently requires extended observational baselines, making it less efficient than other detection techniques. Nevertheless, low-mass stars such as CM Dra represent particularly favourable environments for CBP. Their smaller radii and lower luminosities significantly enhance the detectability of planetary transits and improve the signal-to-noise ratio in radial velocity measurements, thereby increasing the likelihood of confirming planetary candidates in such systems \citep{1998A&A...338..479D,2000A&A...358L...5D,2008A&A...480..563D,2000ApJ...535..338D,2009ApJ...691.1400M}.
Indeed, the high orbital inclination of CM Dra makes it a particularly suitable candidate for the transit method of exoplanet detection  \citep[see][for details]{1998A&A...338..479D,2000ApJ...535..338D}. The first long-term observations of CM Dra for planetary transit detection were conducted by \cite{1998A&A...338..479D} using precise timing measurements to search for small dips in brightness due to a transiting companion \citep{1952Obs....72..199S,1997A&AT...13..233D}. The same technique has been instrumental in confirming planetary-mass companions in other systems, most notably in the case of the pulsar PSR 1257+12, where the existence of an exoplanet was first revealed through transit timing variations \citep{1992Natur.355..145W}. Given its well-characterised eclipsing nature, strong magnetic activity, and the potential presence of a CBP, CM Dra represents a crucial test case for understanding planetary formation in multiple star systems.

The existence of CBPs has been validated through the utilization of both transit and radial velocity methods. Significant contributions to this field have been made by missions such as Kepler and TESS \citep[e.g.,][]{2014MNRAS.444.1873A,2018haex.bookE.156M,Martin2019}. Conversely, the detection of CBPs using ETVs has proven to be more controversial, with one of the most reliable cases being provided by \cite{Goldberg2024} based on dynamical ETVs. However, the light time effect method employed in our study is less definitive, and the presence of stellar activity remains a competing explanation. In the context of post-common envelope binaries (PCEBs), numerous proposed CBPs have been subject to scrutiny, as evidenced by studies such as \cite{Bear2014,Zorotovic2013}.  These studies have disproven certain ETV-based brown dwarf candidates, underscoring the need for rigorous examination. Furthermore, systems like QS Vir have been demonstrated to be dynamically unstable, as reported in \cite{Hardy2015}. While ETV-based detections are valuable, they require careful consideration of alternative explanations, and our CM Dra analysis, although promising, remains cautious without further high-precision observations.

Thanks to satellite telescopes and accurate radial velocities, CBPs \citep{2014MNRAS.444.1873A,2018haex.bookE.156M,Martin2019,Sairam2024} are now more likely to be detected than in previous years, but planets with much longer orbital periods are technically relatively more difficult to find. With the astrometric and imaging capabilities of Gaia and James Webb Space Telescope (JWST) \citep{2006SSRv..123..485G}, and future projects such as PLATO \citep{2014ExA....38..249R}, we may be able to better detect these CBPs. However, Gaia’s sensitivity is optimized for planets with masses of multiple Jupiter masses and orbital separations of a few au \citep{Sahlman2015}. Given these constraints, while Gaia may not directly detect the proposed CBP candidate in CM Dra, long-term astrometric monitoring could still provide valuable constraints on its orbital properties.

Unlike observational methods such as radial velocity measurements, imaging, and microlensing, the ETV method is also an effective technique for detecting circumbinary planets. However, ETV signals can also arise from stellar activity, necessitating a careful analysis of alternative explanations. In particular, M-dwarfs such as CM Dra are known for their strong magnetic activity, which can manifest as starspots, flares, and cyclic magnetic variations. These effects can distort eclipse timing measurements by introducing surface inhomogeneities, affecting the transit duration and centrality. Previous studies \citep{Tran2013,2020MNRAS.491.1820L} have shown that such activity-induced timing shifts can mimic the signal expected from a circumbinary planet, making it crucial to disentangle these effects in any ETV analysis. In this study, we investigate the nature of the observed ETVs in CM Dra and assess whether they are best explained by a circumbinary planet or by stellar activity, such as magnetic cycles and surface spots.

\section{New observations and data analysis}
\label{sec:obs}

\begin{figure*}
\centering
\includegraphics[width=0.95\linewidth]{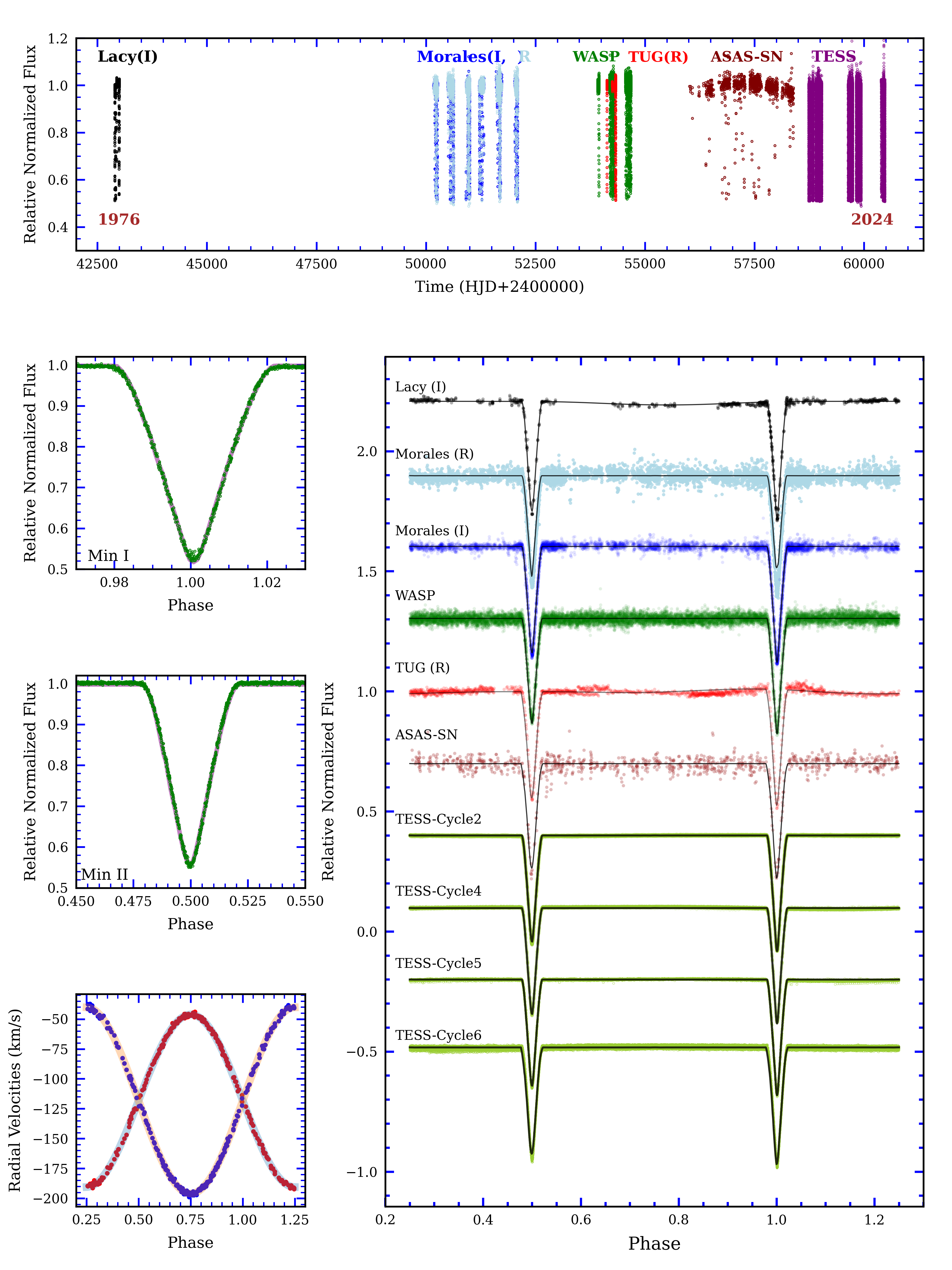}
\vspace*{-3mm}
\caption{The observed and the computed (solid line) light curves (a) and radial velocities (b) of CM Dra. \textit{See} text for details.}
\label{fig:CMDRa:LC}
\end{figure*}

The data analysed in this study were performed using a total of about a million observation points at different wavelengths spread over a period of a half century. During the analysis of the system, ground-based observations obtained by WASP \citep{WASP2006PASP..118.1407P} and ASAS-SN \citep{ASASSN2017PASP..129j4502K} were used in addition to observations previously obtained using ground-based telescopes (\citep{1977ApJ...218..444L,2009ApJ...691.1400M}).  All available light curves of CM Dra are shown in Fig.~\ref{fig:CMDRa:LC} including our newly obtained. 

New ground-based observations of CM Dra were obtained in nine nights in 2007 with \textsc{ANDOR CCD } on the 1.5-m Russian-Turkish telescope (RTT150) at the T\"UB\.ITAK National Observatory (TUG). The system was observed using the Cousin R filter with an exposure time of 30-60 seconds. In addition, to obtain new minima times, the system is observed on 11 nights in 2012, 2013, and 2014 at the Ege University Observatory (EUO) with a 40-cm telescope with the R filter. The reduction and analysis of the frames have been performed by using standard packages \texttt{IRAF} \citep{IRAF1986SPIE..627..733T} to subtract the bias and dark and divide the flat field followed by aperture photometry (\texttt{APPHOT}). The total of four new times of minima lights were obtained by applying a parabolic fit in the Heliocentric Julian Date (HJD) format, then transformed to the Barycentric Julian Date (BJD) format as 54132.50119(13), 54325.29635(4), 54328.46834(5), and 56348.37838(8).
Phases are calculated using the ephemeris corrected in this study (Table~\ref{tab:results}).
The model light curve (solid line) is also shown in Fig.~\ref{fig:CMDRa:LC} (see Sec.~\ref{sec:lcmodel} for details).

\begin{table*}
\begin{center}
\caption{Basic information on photometric observations obtained for CM Dra. Qi in the table indicates the quality of the data sets and N indicates the total number of observation points. Qi values are specified as 1, 2 and 3 (the most precise datasets), and weights are determined accordingly during the solution. JDS and time span denote the JD start time and the number of days between the beginning and the last observation point, in units of days, respectively.} \label{tab:obslog}
\begin{tabular}{lllllll}
\hline
Source                     & Filter(s)  & JDS   & Time span &   Qi   & N      & References    \\
\hline
Lacy                       & I          &42893.8    & 103   & 2    &839     & \cite{1977ApJ...218..444L} \\
Morales                    & RI         &50173.0    & 1917  & 2    &11364   & \cite{2009ApJ...691.1400M} \\
WASP                       & SW         &53920.4    & 761   & 3    &14680   & This study   \\
TUG                        & R          &54132.4    & 196   & 2    &1493    & This study  \\
ASAS-SN                    & V          &56018.1    & 2369  & 3    &976     & This study, \cite{ASASSN2017PASP..129j4502K}   \\
TESS                       & TESS          &58738.7    & 1254  & 1    &235138  & This study, \cite{TESS2015JATIS...1a4003R}   \\
\hline
\end{tabular}
\end{center}
\end{table*}

\begin{figure}
\centering
\includegraphics[width=0.97\linewidth]{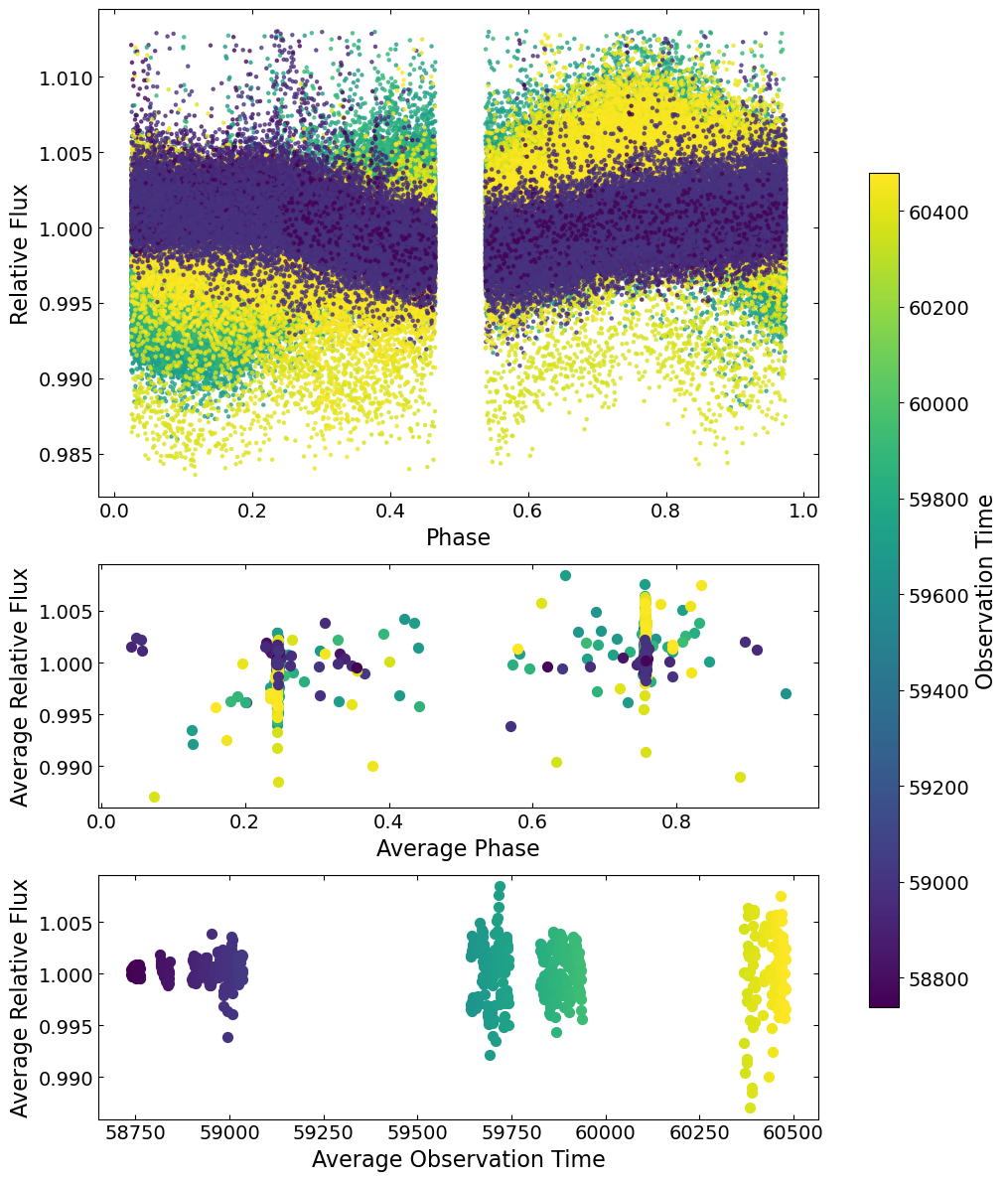}
\caption{Out-of-eclipse variations of the light curves from TESS observations. The colour bar is plotted as a function of time. The top panel shows the variation by phase, while the middle panel shows the mean phase and mean flux for each light curve. The lower panel shows the mean time and mean flux for each light curve.}
\label{fig:CMDRa:LC_Active}
\end{figure}

In addition to ground-based observations, CM Dra was observed by a space-based TESS telescope \citep{TESS2015JATIS...1a4003R}. During the mission, TESS will have run a total of five separate observing campaigns, referred to as Cycles. CM Dra was observed in four Cycles (2, 4, 5 and 6) for 19 Sectors. During "Cycle 2", 111484 observation points spread over a total of 296 days, 49929 observation points spread over 102 days in "Cycle 4" and finally 68351 observation points spread over 111 days in "Cycle 5" were received. These numbers represent the 30-minute exposure times of the TESS observations. In some of the TESS sectors, observations were also made with two-minute exposure times for some selected systems. Short cadence (2 minutes) observations of CM Dra were made in Sectors 49, 50, 51, 52, 56, 57, 58, 59, 76 and 77; 20 seconds in Sectors 76, 77, 78 and 79. At the end of the observations with ground-based telescopes and TESS, $\sim 10^6$ observation points were obtained for CM Dra.

In the analyses of this study, in addition to the observations we have obtained (TUG, EUO), the light curves obtained by the data sets \cite{1977ApJ...218..444L}, \cite{2009ApJ...691.1400M}, ASAS-SN, WASP, and TESS are given together in Figure~\ref{fig:CMDRa:LC}. The observational logs for these data sets are summarized in Table~\ref{tab:obslog}.
The top panel of Figure~\ref{fig:CMDRa:LC} shows the normalised flux versus time. 
In the bottom-left panel of the figure, the minima of the TESS light curve and the radial velocity curves of the system are plotted.
In the bottom right panel, the phase diagrams are plotted for each data set. We downloaded the observation data sets obtained by TESS from the Mikulski Archive for Space Telescopes (MAST) servers and performed the required reduction and detrending processes using the \texttt{Lightkurve} \citep{Lightkurve2018ascl.soft12013L} code. \texttt{PDCSAP} fluxes were used during data reductions and normalisation. \texttt{PDCSAP} fluxes are derived from the simple aperture photometry (\texttt{SAP}) fluxes and are obtained using common trend basis vectors (\texttt{CBV}) of possible trends.
There is a large minimum time limit for the observations. To obtain more accurate minimum times, we folded each of the four light curves into one light curve for long-cadense and calculated one minimum time for short-cadence TESS data. We used a short-cadence dataset to find minimum times for sectors 78 and 79, and for each light curve, we set a minimum time with a weight of unity. We chose to use this method to achieve a more sensitive minimum time. Using TESS data, we obtained 160 times the minimum light. During the period change analysis, we weighted the minimum times obtained from the TESS long-cadence datasets four times more than those obtained in short-cadence and previous cases (Table~\ref{tab:mintimes}).
CM Dra is an active close binary system, and some flare activities have been noted in the literature (see Sections~\ref{sec:intro} and \ref{sec:activity}). In addition to obtaining perfect light curves by TESS, hundreds of flare activities caused by the components of the CM Dra system were also detected.

\section{Stellar activity and flares}
\label{sec:activity}
Space-based telescopes such as Kepler and TESS, which can obtain precise photometric variations and have continuous observation series, allow us to study both exoplanets and variable stars well and to observe flaring stars in detail. Especially active stars of the M-type can frequently exhibit flares, superflares, and two flares \cite{Gunther2020AJ....159...60G,Ilin2021MNRAS.507.1723I}. Due to the stellar spot activities, the light curve of CM Dra is observed to change (see \cite{2009ApJ...691.1400M}). Outside the eclipses, the brightness variation is of the order of $\sim$0.02 mag. A variation in the phase of the maximum brightness of CM Dra from 0.28 to 0.60 of the orbital period between 1977-1997 is explained with an asymmetric large, long-lived polar spot reported \citep{2004ARep...48..751K}. 

One of the best methods to follow stellar surface activity is the use of consecutive photometric light curves. Since there are significant gaps of days between the light curves obtained in previous years of the CM Dra eclipsing binary system, it is better to use only successive TESS observations for this purpose. The light curves of all TESS sectors are combined and eclipses are subtracted to show the normalised flux variation versus phase in the top panel of Fig.\ref{fig:CMDRa:LC_Active}. All flares, i.e. all sudden brightness variations, are also subtracted in this plot. Thus, the change in the average total surface brightness is shown. As can be seen in the figure, significant brightness variations occur at the maximum phases with time. Moreover, the brightest parts at the first maximum (around phase 0.25) are reduced to a minimum at later times and the opposite is observed at the second maximum (around phase 0.75). This variation shows that the stellar surfaces change over short periods of time.  The middle panel of the same figure shows the average flux variation versus the average phase. In the bottom panel, the mean flux versus mean time is also plotted. These plots show that CM Dra was less active in the first part of the TESS observations and more active in the later observations. These figures also show that we should include spot models in light curve analyses and that non-conservative mass losses are necessary in stellar evolution models.

Although CM Dra is known to be active and regularly flares from observations made for years, TESS observations clearly show flare numbers and flare properties. 
It has been previously reported that the component stars of CM Dra exhibit activity and have shown various numbers of flares (see \cite{2007IBVS.5789....1N} and references therein).

\begin{figure*}
\centering
\includegraphics[width=0.8\linewidth]{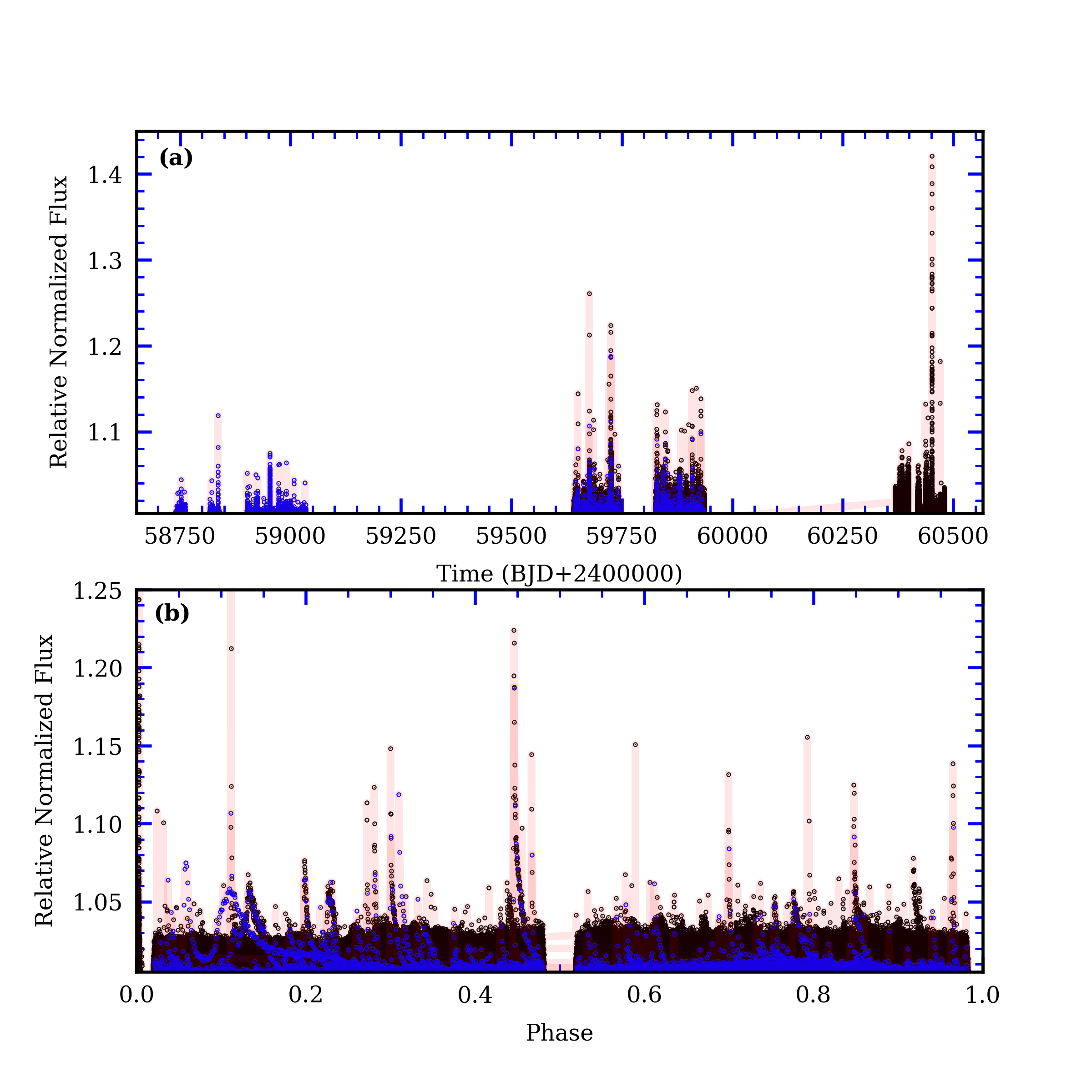}\vspace{-0.4cm}
\includegraphics[width=0.8\linewidth]{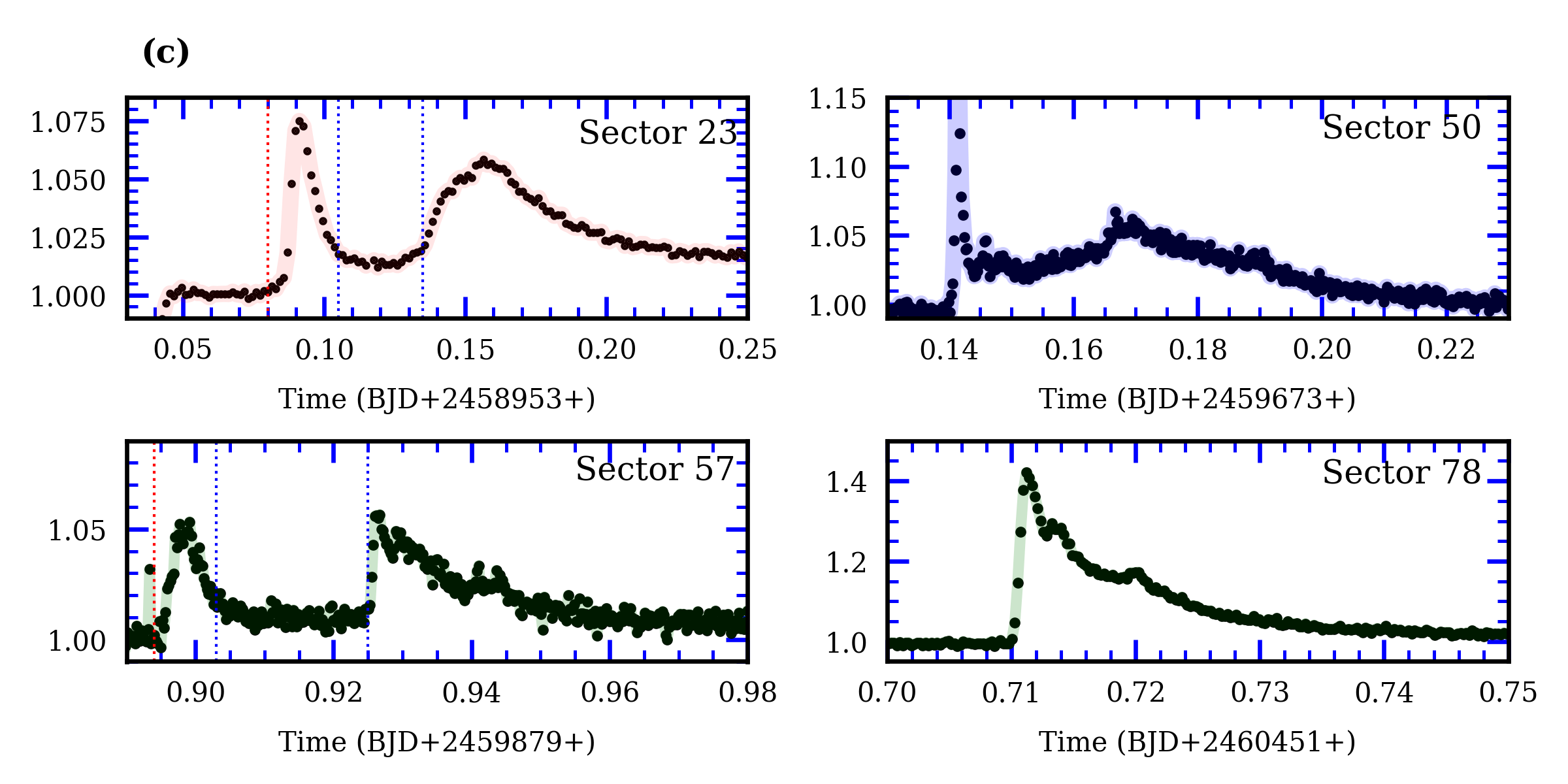}
\caption{Time (a) and phase (b) distribution of flares seen in the CM Dra system in TESS observation sets. In (a) and (b), both the 2-minute (blue) and 30-minute (black) exposure times are superimposed. Four flare patterns showing different characteristics are shown in (c).}
\label{fig:cmdra_flares}
\end{figure*}

The light variations obtained from 19 TESS Sectors (16, 19, 22, 23, 24, 25, 26, 49, 50, 51, 52, 56, 57, 58, 59, 76, 77, 78 and 79) are plotted in Figure~\ref{fig:cmdra_flares}, with the variation of the out-of-eclipse flux zoomed in to show the flares in the time domain (BJD) and folded light curve. In each sector of the TESS observations of CM Dra, large flare observations were observed. Recently, flaring activity in CM Dra has been studied in detail by \cite{Martin2024}. In their study, they have very accurately studied all flare activities in the CM Dra system excluding the last four sectors of TESS. We will not reproduce here the detailed flare study done by them. They reported 163 flares using TESS data sets and found the activity cycle with 4.1 yr.

After the \cite{Martin2024} study was published, the TESS Sector 76, 77, 78 and 79 observations were published and a large number of flares were also observed in these observations (see right side of Figure \ref{fig:cmdra_flares}a). It is also worth noting that flares have been detected not only during the maximum phases but also sometimes when approaching and passing through eclipses, and sometimes during both the primary and secondary eclipses. This suggests that both stellar components exhibit magnetic activity. Another issue is that a total of $\sim 4.8$ years of TESS data of the system shows that the flare activity in Maximum I is higher than the activity in Maximum II (Figure~\ref{fig:cmdra_flares}b).

In addition to flare observations with a single peak, we detected consecutive flares with two peaks observed in the system. We have highlighted these flares and show the most prominent ones separately at the bottom of Figure~\ref{fig:cmdra_flares}c). In Sectors 49, 50 and 57, data sets with exposure times of 2 minutes were used. In Sectors 23, 24 and 25, data with exposure times of 30 minutes were used because there were no short-cadence data in the TESS Sectors earlier than 27. Such flares can provide us with information about either successive outbursts, events near the poles, or possible flare eclipses. 
An interesting phenomenon has been detected during TESS observations of CM Dra (e.g., Fig.~\ref{fig:cmdra_flares}c), Sectors 23, 50 and 57). A flare is detected and, after the maximum phase of the flare, a planet transit is detected, or maybe a flare is eclipsed by one of the components of the system with the related inclination. These phenomena are very rare and may have been detected in this study. However, we note that such an event is rare and that there may be another possible explanation for these changes. The transit event took about 9.4 and 4.9 hours, and the transit time is about 44 and 30 minutes for the flare observed in Sectors 23 and 57 respectively. The beginning and ending points of these periods are indicated by the red line in Fig.~\ref{fig:cmdra_flares}c.

\section{Analysis of eclipse timings}
\label{sec:pca}
The time variation of the difference between the observed (O) and calculated (C) minimum times of a lattice binary system gives us clues about the processes that cause the system to change orbital period. The shape of the $O-C$ diagram of an eclipsing binary gives information about the possible mass transfer between the components and/or any third body in the system. Sinusoidal modulation in the $O-C$ values indicates the presence of a third body in a binary system. The $O-C$ variation of the CM Dra eclipsing system has been studied in detail by \cite{2008A&A...480..563D}. The authors reported the possible presence of a third body orbiting CM Dra and proposed a few Jupiter mass (M$_{\textrm{J}}$) object with 18$\pm$4.5 yr orbital period or an object in the mass range of 1.5 M$_{\textrm{J}}$ to 0.1 M$_{\odot}$ with orbital periods that extend to thousands of years, as a third body candidate.

\begin{figure*}
\centering
	\includegraphics[width=0.92\linewidth]{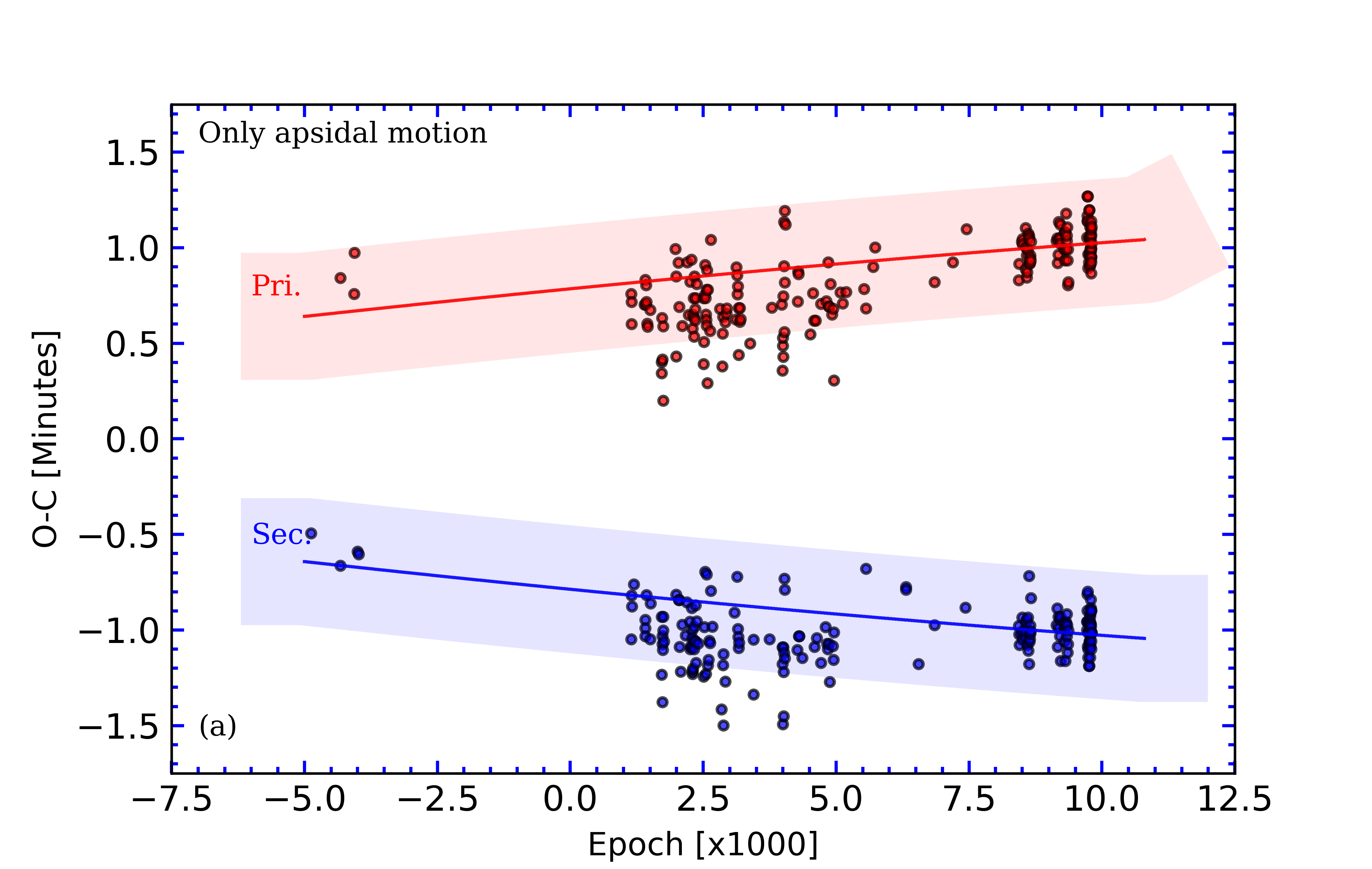} \\ 
 \vspace{-7.0mm}
	\includegraphics[width=0.92\linewidth]{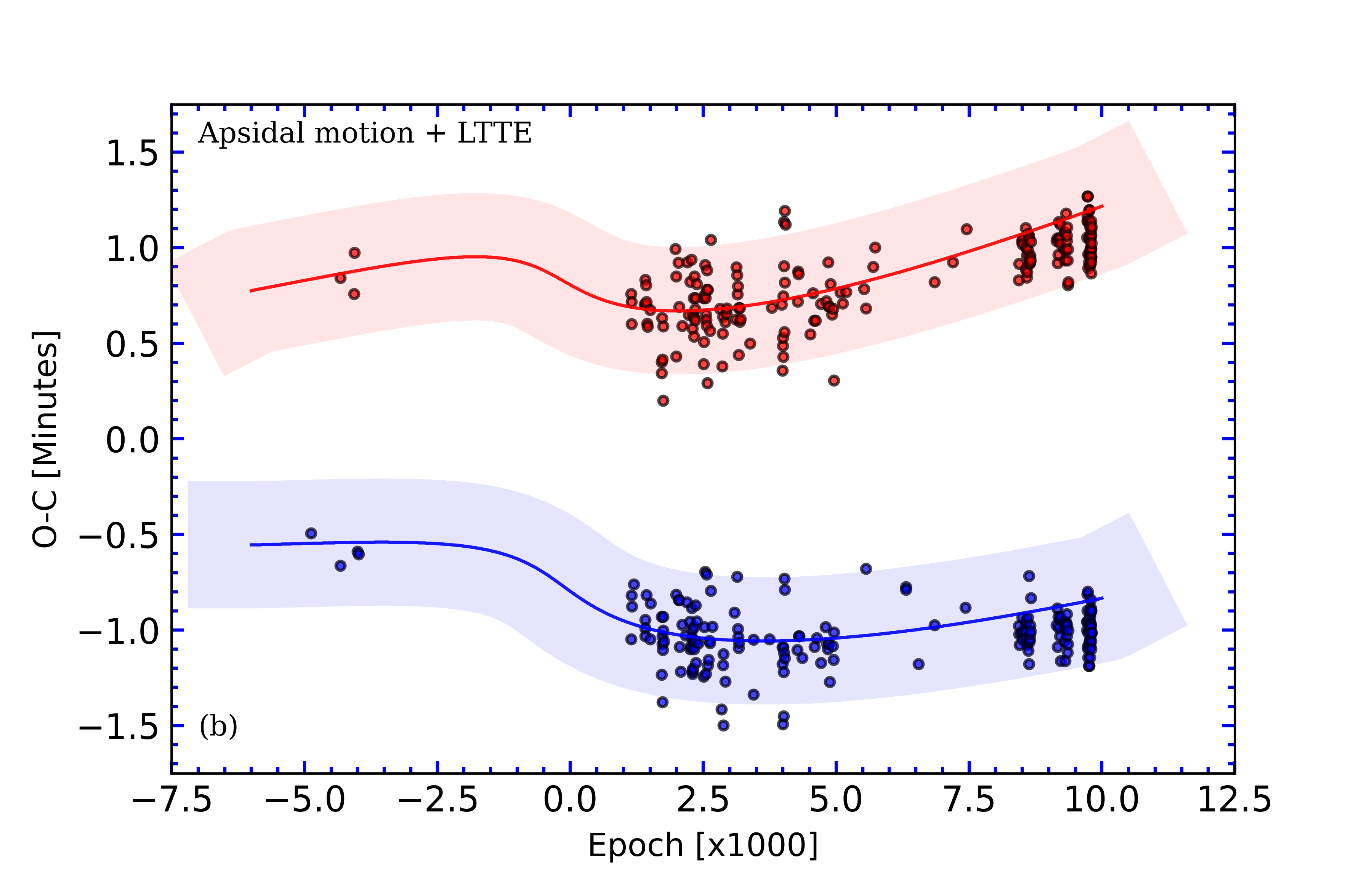} 
\caption{The primary (red) and secondary (blue) times of minimum light of CM Dra. The result of the analysis of the period change is based only on the effect of apsidal motion (upper panel) and the variations based on the effect of a possible planet in addition to the apsidal motion (lower panel).}
\label{fig:CMDra:OC}
\end{figure*}

CM Dra has been studied extensively and numerous instances of minima exist in the literature.
The times of minima light obtained until 2009 were given by \cite{2009ApJ...691.1400M}.
In addition to those data, period variation analysis is performed using the data given by \cite{2009IBVS.5898....1P,2011IBVS.5980....1P,2014IBVS.6114....1Z}  and one is from the BRNO website.
We have collected 307 times of minima for period-change analysis. 
A total of 164 new times of minima obtained in this study, together with all the minima times used in the study, are given in Table~\ref{tab:mintimes}. We have shown the O-C changes of all the minimum times obtained in Fig.~\ref{fig:CMDra:OC}.  We have collected data from many different instruments with different filters. This will of course contribute to some of the scatter seen in Fig.~\ref{fig:CMDra:OC}. Of course, this does not produce a change that causes sinusoidal fluctuations and only partially scattered data sets.

Apsidal motion can be determined by long-term observations of the changing positions of eclipses in the light curve.  Generally, the changes due to apsidal motion result from the nonhomogeneous interior structure of the component stars and the interaction between stars with eccentric orbits and rotational distortions.
For the period change analysis showing apsidal motion, the times of mideclipses listed in Table~\ref{tab:mintimes} were converted to BJD.
The analysis is done with the weighted least-squares method. The equations (Eqns.~\ref{eq1}-\ref{eq4}) for the case of a third body, apsidal motion and explanations are taken from \cite{1983Ap&SS..92..203G,1995Ap&SS.226...99G,2009ApJ...691.1400M}. The relation between apsidal motion and binary parameters is given in Eqs.~\ref{eq2}-\ref{eq4:rel}.
In our analyses to determine orbital parameters ($T_0$, $P_{\rm s}$, $e$, $\dot{\omega}$, and $\omega_0$), we used the code developed by \cite{Zasche2009NewA...14..121Z} and included eccentricity terms up to the fifth order in the expressions given by Equation~\ref{eq1}.
We have written only the linear term of Equation 1 and denoted the other four terms by $\chi(e)$ since the subsequent terms have a much longer notation. The explicit expression of the other terms is found in the study of \cite{1995Ap&SS.226...99G}.

The parameters in Equation~\ref{eq1} are $E$, $A_1$, $\omega$, $\dot{\omega}$, $P_s$, $P_a$, and $U$, respectively, and these terms represent the orbital cycle, a coefficient depending on the inclination and orbital eccentricity, the argument of periastron, the rate of apsidal motion, sidereal period, anomalistic period, and apsidal motion period, respectively. $n$ is for the minimum times of primary ($n = 1$) and secondary ($n = 2$) of the light curves. Please, see references \cite{1995Ap&SS.226...99G, 2008A&A...480..563D,2009ApJ...691.1400M} for exact expressions and descriptions of the equations.

\begin{table}
\begin{center}
\caption{Times of primary and secondary minima for CM Dra in BJD (days). 
'P' and 'S' indicate whether the minimum corresponds to the primary or secondary eclipse. All the data of this table is published in its entirety in the electronic edition of the Journal. }
\label{tab:mintimes}
\begin{tabular}{llllll}
\hline
BJD	        &   P/S	& 	BJD	        &   P/S	&    BJD	    &   P/S	 \\
\hline
2441855.75510	&   S	& 	2442966.86474	& 	S	& 	2449500.97580	& 	P   \\
2442555.90638	& 	S	& 	2442994.76932	& 	S	& 	2449501.61000	& 	S   \\
2442557.80896	& 	P	& 	2449494.63385	& 	P	& 	2449511.75712	& 	S   \\
2442888.85875	& 	P	& 	2449497.80483	& 	S	& 	2449562.49272	& 	S   \\
2442893.93231	& 	P	& 	2449499.70741	&   P	& 	2449815.53652	& 	P   \\ 
\hline
\end{tabular}
\end{center}
\end{table}

\begin{multline}\label{eq1}
T_n \simeq T_0+P\left(E+\frac{n-1}{2}\right) + (2 n-3) A_1 \frac{e P}{2 \pi} \\
\times \left(\cos \omega_0-\sin \omega_0 \cdot \dot{\omega} E\right) + \chi(e)
\end{multline}

\begin{equation}
\omega = \omega_0 + \dot{\omega}\times E \label{eq2}
\end{equation}

\begin{equation}
P_s =P_a (1-\dot{\omega}/360^\circ) \label{eq3}
\end{equation}

\begin{equation}
U= 360^\circ(P_a/\dot{\omega}) \label{eq4}
\end{equation}

\begin{equation}
\dot{\omega}_{\text {rel }}=5.447 \times 10^{-4} \frac{\left(M_1+M_2\right)^{2 / 3}}{\left(1-e^2\right) P_{\mathrm{a}}^{2 / 3}}
\label{eq4:rel}
\end{equation}
  
The light variation analysis of CM Dra by \cite{2009ApJ...691.1400M} does not show any significant  rate.
The authors, on the other hand, used the separation between primary and secondary times of minimum to study any possible  and reported a clue for the presence of  in CM Dra.
Observed and calculated times of minimum light span over a half-century are shown in Fig.~\ref{fig:CMDra:OC}.
The location of the primary and secondary minima in Fig.~\ref{fig:CMDra:OC}a is a clear indication of an .
In Fig.~\ref{fig:CMDra:OC} the computed residuals $O-C$ are obtained using the linear parameters $T_{\rm 0}=24~48042.32731(6)$ and $P=1.26838999(2)$. Fig.~\ref{fig:CMDra:OC}b shows the orbital period variation assuming the possible existence of a third body. This variation is due to the third-body orbiting a close binary system CM Dra.

Table~\ref{tab:apsidal} lists the parameters of the third body with the orbital parameters of the binary system obtained from the analysis. The results indicate that the mass function of the third body is $f(m)=8.7\times10^{-9}$ M$_{\odot}$ with an orbital period of 56 years. The angle between the orbital plane of the third body and the sky plane ($i_P$) determines the mass of this object orbiting CM Dra to be 1.2 $\rm{M_J}$, 1.4 $\rm{M_J}$, and 2.4 $\rm{M_J}$ for 90$^\circ$, 60$^\circ$, and 30$^\circ$; respectively. Long-term variations derived from relatively short-term observations have large uncertainties, as expected. The uncertainty in the mass of the possible third object of the CM Dra system was calculated to be 38\%. The mass values and errors for different $i_P$ angles under this uncertainty are given in Table~\ref{tab:apsidal}. The uncertainty in the period is slightly higher and amounts to 68\%. This corresponds to an error of 38.1 years in the period. To obtain a smaller error, at least one more cycle of observations is needed, i.e. another half century of observations. Despite the uncertainty, the result provides a plausible estimate of the companion’s orbital period. We found that the period of the  is out 2670 years. The time interval of the data set that we use to find such a long period of  is only 51 yr. This is a significant obstacle in the calculation of the error. However, \cite{2009ApJ...691.1400M} gave the uncertainty of the  period as \%59. With a similar uncertainty, the error in this study is around 1170 years, which is quite large.  

Our analysis indicates that the eclipse timing variations observed in CM Dra can be attributed either to the light-time effect from a long-period planetary companion or to magnetic activity cycles. Previous studies by \citet{2013ApJ...768...33R} and \citet{Tran2013} demonstrated that quasi-periodic ETVs in close binaries observed by \emph{Kepler}, with timescales of 50–200 days and amplitudes of 200–300 seconds, can be explained by magnetic activity and spot modulation, often displaying anticorrelated timing shifts between primary and secondary eclipses. Given the known magnetic activity of CM Dra, such effects could plausibly contribute to its $O-C$ variations. While a third object with a planetary mass and an orbital period of about 56 years provides a statistically improved fit to the ETVs, the current data do not decisively favour this over magnetic mechanisms. Therefore, both scenarios remain viable and cannot yet be distinguished conclusively.

To further assess the relative quality of the apsidal motion (AM) and AM + LTTE  (light-travel time effect) models, we computed the Bayesian Information Criterion (BIC) for each solution. In this calculation, the number of data points ($n$) corresponds to the number of eclipse timing measurements used, not the number of individual photometric data points. Using the sum of squared residuals (SSR), number of fitted parameters ($k$), and the total number of data points, we found that the AM + LTTE model yields a significantly lower BIC value ($-6831.4$) compared to the AM-only model ($-6792.8$). The resulting $\Delta$BIC provides statistical support for the AM + LTTE model \citep{schwarz1978,kass1995}. Nonetheless, this statistical preference does not conclusively confirm the presence of a third body, as stellar activity and other sources of red noise may also contribute to the observed variations. Starspots and/or flares on active stars such as CM Dra, especially M dwarfs, can cause asymmetries in the minima of the light curve due to their total or partial eclipses during the transit, affecting the ETVs calculations. Therefore, we would like to stress that both cases are possible and only new observations in the future will be able to reveal this situation more reliably. Therefore, although the BIC analysis statistically favours the AM + LTTE model, stellar activity presents a credible alternative that warrants consideration until further observational evidence becomes available. 
While the eccentricity values obtained from the AM-only, AM+LTTE, and the simultaneous LC+RV solutions are not identical, they are similar in magnitude. The AM+LTTE solution yields $e = 0.008 \pm 0.001$, while the RV-only fit gives $e = 0.005 \pm 0.001$, and the simultaneous LC+RV model yields $e = 0.004 \pm 0.001$. These differences likely reflect the differing nature of the datasets and modelling sensitivities involved in each approach. In particular, the ability to resolve such small eccentricities in the AM+LTTE framework is made possible by the precision and time span of the new datasets.

Although the AM+LTTE model provides a better statistical fit to the O–C variations, it is important to note that stellar magnetic activity, particularly flares, can significantly distort eclipse profiles and alter the derived eclipse timings. Many flares were detected across multiple TESS sectors \citep[e.g.,][]{Martin2024,Armitage2025}, including events with complex, multi-peaked morphologies, some of which occurred during or near eclipse phases. These flares may bias individual timing measurements, particularly when combined with varying activity levels over time and across observing passbands. While the LTTE model remains viable, this complexity introduces an additional layer of uncertainty that must be considered when interpreting ETVs. Future work is needed to quantify the impact of flare morphology and cadence-dependent light curve distortion on timing precision.

\begin{table}
	\begin{center}
 \scriptsize
		\caption{Apsidal motion (AM) and light-travel time effect (LTTE) elements for CM Dra.}\footnotesize \label{tab:apsidal}
		\begin{tabular}{lll}
			\hline
			Parameter                                    &  AM+LTTE   &  AM        \\
			\hline
			\textit{\textbf{Binary star orbit }}         &                   &           \\
			Epoch, T$_{\rm o}$   (day)                   &448042.32731(6)    & 448042.32716(7)\\
			Period, P  (day)                             & 1.26838999(2)     & 1.26839001(2)  \\
			Eccentricity, e                              & 0.0082(6)         & 0.0022(3)         \\
			Argument of periapse, $\omega$${({^\circ})}$ & 99(2)             & 129(2)       \\
            d$\omega / dt$  (deg/cycle)                  & 0.0003(1)         & 0.0016(10)      \\
			Period, P$_a$  (day)                         & 1.26839097(10)    & 1.26839562(9)  \\
			Apsidal motion period, U(yr)                 & 4541(1170)        & 786         \\
            Semimajor axis, $a_3$ (AU)                   & 0.028(12)         &              \\
            Periastron passage (HJD)                     & 2448069           &              \\
			Semiamplitude of the LTTE(d)                 & 0.00013(4)        &               \\
            Orbital period, $P_{\rm 3}$(yr)              & 56(38)            &                \\
			Eccentricity, $e_{\rm 3}$                    & 0.59              &                \\
			Minimum Mass, $M_3$ ($\rm{M_J}$)             & 1.2(4)            &                \\
			Mass ($i_P=60^{\rm o}$), $M_3$ ($\rm{M_J}$)  & 1.4(5)            &                \\
			Mass ($i_P=30^{\rm o}$), $M_3$ ($\rm{M_J}$)  & 2.4(9)            &                \\
			\hline
		\end{tabular}
	\end{center}
\end{table}

\begin{table}
	\begin{center}
		\caption{Parameters for CM Dra (A, B) based on simultaneous solution. The standard errors 1$\sigma$ in the last digit are given in parentheses.} \label{tab:results}
		\begin{tabular}{llllll}
			\hline
			Parameter                                           &  Value             \\
			\hline
			\textit{\textbf{Binary star orbit }}                & \\
			Initial epoch, T$_{\rm o}$   (day)                  & 58738.65954(15) \\
			Period, P  (day)                                    & 1.268390011(3)    \\
			Inclination, i ${({^\circ})}$                       & 89.96(6)        \\
			Eccentricity, e                                     & 0.0038(1)        \\
			Argument of periapse, $\omega$ ${({^\circ})}$       & 119(3)           \\
			Semi-major axis length, $a$ ($\rm{R_{\odot}}$ )     & 3.7612(53)       \\
			Distance from Earth, d (pc)                         & 14.4(6)           \\
			\textit{\textbf{Star A}}                            &                     \\
			Mass, $M_A$ ($\rm{M_{\odot}}$)                      & 0.2307(8)            \\
			Radius, $R_{\rm A}$  ($\rm{R{\odot}}$)              & 0.2638(11)             \\
			Effective temperature, T$_A$  (K)                   & 3130(70)                   \\
			Luminosity, $L_A$ ($\rm{L_{\odot}} $ )              & 0.0060(5)               \\
			Surface gravity, $\log g_A$ (cgs)                   & 4.959(6)               \\
			Absolute bolometric magnitude, M$_{\rm b}$  (m)   & 10.29(10)                 \\
			\textit{\textbf{Star B}}                            &                            \\
			Mass, $M_B$ ($\rm{M_{\odot}}$)                      & 0.2136(8)                 \\
			Radius, $R_{\rm B}$  ($\rm{R_{\odot}}$)             & 0.2458(10)                 \\
			Effective temperature, T$_{\rm B}$  (K)             & 3103(70)                  \\
			Luminosity, $L_B$ ($\rm{L_{\odot}} $ )              & 0.0050(4)                 \\
			Surface gravity, $\log g_{\rm B}$ (cgs)             & 4.987(6)                 \\
			Absolute bolometric magnitude, M$_{\rm b}$  (m)   & 10.49(9)                  \\
			\hline
		\end{tabular}
	\end{center}
\end{table}

\section{Light curves modelling and astrophysical parameters of the system}
\label{sec:lcmodel}

The newly obtained light variation of CM Dra has been combined with the light and radial velocity curves published in the literature \citep{1977ApJ...218..444L,1996ApJ...456..356M,2009ApJ...691.1400M}. For the analysis of light and radial velocity curves of the CM Dra, we used the Phoebe code \citep{2005ApJ...628..426P} and the Wilson-Devinney (W-D) \citep{wilson1971ApJ...166..605W, wilson1979ApJ...234.1054W} code based on differential corrections (DC), including a Levenberg-Marquardt scheme (LM). The code W-D has been very successfully tested over a half-century in determining parameters of eclipsing binary stars. This code is also very capable of solving light curves, including stellar spots. Since TESS observations, especially short-cadense data sets, are more sensitive than ground-based telescopes, in the light-curve analyses, we first combined the TESS light curves with the radial velocity curves and obtained the precise parameters. Then, assuming that the basic parameters of the star (mass, radius, etc.) did not vary in the short term, we modelled the light curves of the observations obtained from other ground-based telescopes by only varying the spot parameters.

The results are summarized in Table~\ref{tab:results}. Where \textit{i} is the orbital inclination; \textit{$\omega$} is the longitude of periastron; \textit{e} is the orbital eccentricity. The fractional radius of the components is obtained as 0.07010(3) and 0.06535(2) for the primary and secondary components, respectively. The light curve analysis of the system indicates a well-detached system. The most accurate physical parameters of stars can be obtained by an eclipsing, detached, double-lined spectroscopic binary system. The orbital and physical parameters of CM Dra are determined very accurately in this study by using a simultaneous solution of all available observations (\textit{see} Table~\ref{tab:results}). 
Gaia gave a parallax of 67.29$\pm$ 0.03, mas \citep{2021A&A...649A...1G} for CM Dra.
The accuracy of the masses, which depends mostly on the accuracy of the light and radial velocity data, is less than 1\%. Using these parameters, we determined the distance of the system to be $46.8\mp0.3$ light years, which is very close to the distance obtained by Gaia as 48.4 light years.

\cite{Martin2024} claimed that they obtained the parameters of the component stars of CM Dra very precisely. Although the method they followed was to use light and radial velocity data sets, the deviations from the fit were unrealistically small because the code they used to model the light curve and the number of parameters they allowed were not as large as necessary. However, the light variations of interacting close binary stars like CM Dra depend on many parameters. Among the most important of these parameters are the orbital inclination angle (i), temperature ratios of the component stars (T$_2$/T$_1$), mass ratios (q), albedos (A$_{1,2}$), potentials ($\Omega_1$, $\Omega_2$), third-body effect, limb- and gravitational darkening and, in CM Dra case, the hot or cool spots effect on the stellar surface. The modelling of light variations due to such a large number of parameters has been described in detail in the \cite{wilson1971ApJ...166..605W, wilson1979ApJ...234.1054W, prsa2005ApJ...628..426P} studies. To give an example, the radius errors obtained without taking surface potentials into account are in the order of tens of thousands, while the radius errors in the analysis with surface potentials are in the order of thousands. 

\cite{2009ApJ...691.1400M} obtained precise physical parameters from the solution of light curves obtained from ground-based observations. In this study, we got more precise parameters because we analysed observations from ground-based and space-based telescopes together. The uncertainties in mass and radius are twice as significant. In addition to the uncertainty, we found the stellar masses of the primary and secondary components to be 0.1 and 0.2 per cent smaller than in \cite{2009ApJ...691.1400M} study and 2.5 and 1.7 per cent larger than in \cite{Martin2024} study. The difference between the radii of the stars is a bit more significant. For the first component, we got about 4 per cent larger and for the second star, about 3 per cent larger.

\section{Results and Conclusion}
\label{sec:results}

Because of the low-mass nature of the components, their nuclear time scales are expected to be longer than Hubble time. The stars, on the other hand, are close and active; therefore, they can evolve in a way that is somewhat different and relatively fast. The presence of a third body around a close binary can cause a periodic variation in the orbital period of a binary system.
A usual method to study this variation is to calculate the variation of the observed times of minimum light. The presence of a third body can be revealed by measuring as many precise times of minimum light as possible. This technique is also one of the methods to detect companions of the planet.

The integrated analysis of ground-based and space-based observations of CM Dra, an active and close-eclipsing binary system with very low-mass components, has yielded significant findings. Firstly, the orbital and physical elements were obtained with great precision through the use of synthetic modelling techniques applied to the double-line radial velocity and light curves, as detailed in Tables~\ref{tab:apsidal} and \ref{tab:results}. The obtained minimum times for the system were subjected to analysis, and a period change analysis was conducted under a range of assumptions. The period variation illustrated in Figure~\ref{fig:CMDra:OC} was initially examined under the assumption of solely apsidal motion. The residual observations (Fig.~\ref{fig:CMDra:OC}b) indicate the presence of a sinusoidal variation superimposed on the variations attributable to apsidal motion. 

To more robustly compare the AM and AM + LTTE models, we computed the Bayesian Information Criterion. This approach accounts not only for the sum of squared residuals (SSR) but also for the number of fitted parameters, providing a balance between goodness of fit and model complexity. The BIC values derived from our analysis are $-6792.8$ for the AM-only model and $-6831.4$ for the AM+LTTE model. The resulting $\Delta \mathrm{BIC}$ of 38.6 suggests that the AM+LTTE model provides a better statistical fit than the AM-only model, although this does not constitute definitive evidence. However, this statistical preference should not be interpreted as definitive evidence for the presence of a planetary companion. While the LTTE hypothesis provides a statistically favourable explanation within the current data limits, alternative mechanisms—particularly stellar magnetic activity—remain plausible contributors to the observed ETVs.

The orbital period change analysis results are detailed in Table~\ref{tab:apsidal}. The period of apsidal motion of CM Dra is found to be $4541$ years. The apsidal motion rate ($\dot{\omega}$) is the sum of the contributions of classical ($\dot{\omega}_{\rm {cl}}$)  and general relativistic  ($\dot{\omega}_{\rm {GR}}$) contribution. The expressions given by Equation \ref{eq4:rel} have been formalised by \cite{levi10.2307/2371404,Kopal1978ASSL...68.....K,Shakura1985SvAL...11..224S,Gimenez1985ApJ...297..405G}. We calculated the relativistic contribution of the apsidal motion rate as $\dot{\omega}_{\rm {GR}}=0.00027$  deg cycle$^{-1}$ and the classical contribution estimated as $\dot{\omega}_{\rm {cl}}=0.00046$ deg~cycle$^{-1}$  for CM Dra. These results show that mutual tides raised within the CM Dra might drive the apsidal motion. The observed period variations may be explained either by the light-time effect of a third body or by stellar magnetic activity, both offering plausible interpretations within the current observational constraints. If real, such an object would likely reside in the planetary mass regime. However, this interpretation relies on the assumption that the variations are not primarily caused by stellar activity. Of course, this was done on the assumption that a third object was the effect causing the change in orbital period.  As we mentioned in Section~\ref{sec:activity}, stellar spots can also affect the $O-C$ variation \citep{1992ApJ...385..621A,2020MNRAS.491.1820L,2010MNRAS.405.2037W}.
We emphasize that stellar activity remains a viable alternative explanation for the observed O-C residuals. Starspots, differential rotation, magnetic cycles, and episodic mass ejections can induce quasi-periodic variations in the timing of eclipses, potentially mimicking the effects of a third body. Observations of similar systems, such as those discussed in \cite{Bear2014, Zorotovic2013, Hardy2015}, have revealed that spot-induced asymmetries in light curves can translate into timing variations, particularly in magnetically active low-mass binaries. Additionally, studies such as \cite{Tran2013} and \cite{2020MNRAS.491.1820L} suggest that magnetic braking and angular momentum redistribution due to surface activity can introduce long-term variations in the orbital period. These mechanisms must be considered when interpreting the observed ETVs in CM Dra, and further spectroscopic and photometric monitoring will be necessary to disentangle planetary-induced variations from stellar activity.

The presence of a third body (e.g., a planet or a star) orbiting a binary system can induce dynamical instabilities, leading to periodic variations in the eccentricity and orbital inclination of the binary components. This phenomenon, known as the von Zeipel-Kozai-Lidov (ZKL) effect \citep{vonZeipel1910AN....183..345V,Kozai1962AJ.....67..591K,Lidov1962P&SS....9..719L}, is particularly relevant in hierarchical systems where an external perturber can drive long-term orbital evolution \citep{Kiseleva1998MNRAS.300..292K,Naoz2016ARA&A..54..441N,Toonen2016,Martin2016MNRAS.455L..46M}. The CM Dra system is a compelling example of a dynamically complex multiple system, potentially hosting both a distant white dwarf companion and a circumbinary planet. Recent astrometric studies suggest the presence of a white dwarf at a projected separation of $\sim$5000–10,000 AU from the binary \citep{Borgman1983,2009ApJ...691.1400M,McCleery2020,Rowan2023}, reinforcing the idea that CM Dra is a rare hierarchical triple system. Although the white dwarf’s direct gravitational influence on the CBP is expected to be minimal, its presence in the system suggests that CM Dra could be a rare example of a hierarchical triple system. This configuration raises questions about the long-term dynamical evolution of the system, requiring further study. The identification of a planetary companion in such a system would provide valuable insights into the formation and dynamical evolution of circumbinary planets in multi-star environments.

Triple-star planetary systems remain relatively rare, and their orbital architectures are still not well understood. The CM Dra system offers a unique opportunity to study the interactions between a low-mass eclipsing binary, a circumbinary planet, and a distant tertiary white dwarf in a potentially stable configuration. Previous studies on hierarchical systems \citep{Munoz2015,Martin2015,Hamers2016,Toonen2016} have demonstrated that an intermediate planetary-mass companion situated between an inner close binary and an outer tertiary can experience long-term eccentricity oscillations and even destabilization. Thus, precise characterization of the white dwarf’s orbit is crucial to assess its impact on the CM Dra system’s architecture and long-term stability. Future high-resolution astrometric and spectroscopic observations will be necessary to refine the orbital properties of both the white dwarf and the CBP, enabling a more complete understanding of the system’s hierarchical dynamics. Additionally, in systems with multiple gravitational perturbers, the quadrupole and octupole ZKL mechanisms can influence the planet’s orbital evolution, potentially altering its eccentricity over time \citep{Lithwick2011,Naoz2016ARA&A..54..441N,Antognini2015MNRAS.452.3610A,Martin2016MNRAS.455L..46M}. Further studies incorporating long-term dynamical simulations will be essential to determine whether the proposed CBP could remain stable over gigayear timescales within this complex stellar environment.

The fact that active and interacting low-mass and very close binary systems lose up to twenty per cent of their initial mass during their evolution has previously been shown by nonconservative evolution models by \cite{Yakut2005ApJ...629.1055Y}. Naturally, if the orbital period were sufficiently large, non-conservative evolution would be much less effective \citep{Yakut2015MNRAS.453.2937Y}. The fact that CM Dra has a very low mass and has a relatively short period of 1.2 days, if not a very short period, should not neglect the fact that it also loses some mass compared to its initial mass with a similar dynamo. \cite{Somers2020ApJ...891...29S} have modelled a sequence evolution model, considering the effect of starspots with solar metal abundances, which they named \texttt{SPOTS}. During these models, grid models were generated according to a parameter defined as the total spot-filling factor (f$_{\rm spot}$). f$_{\rm spot}$ is a parameter that ranges from zero to one and represents the magnetic activity scale of the star. This study also demonstrated that the dynamo significantly affects stellar evolution, as revealed by the models for close binary systems made by \cite{Yakut2005ApJ...629.1055Y}.

\begin{figure*}
\centering
\includegraphics[width=0.45\linewidth]{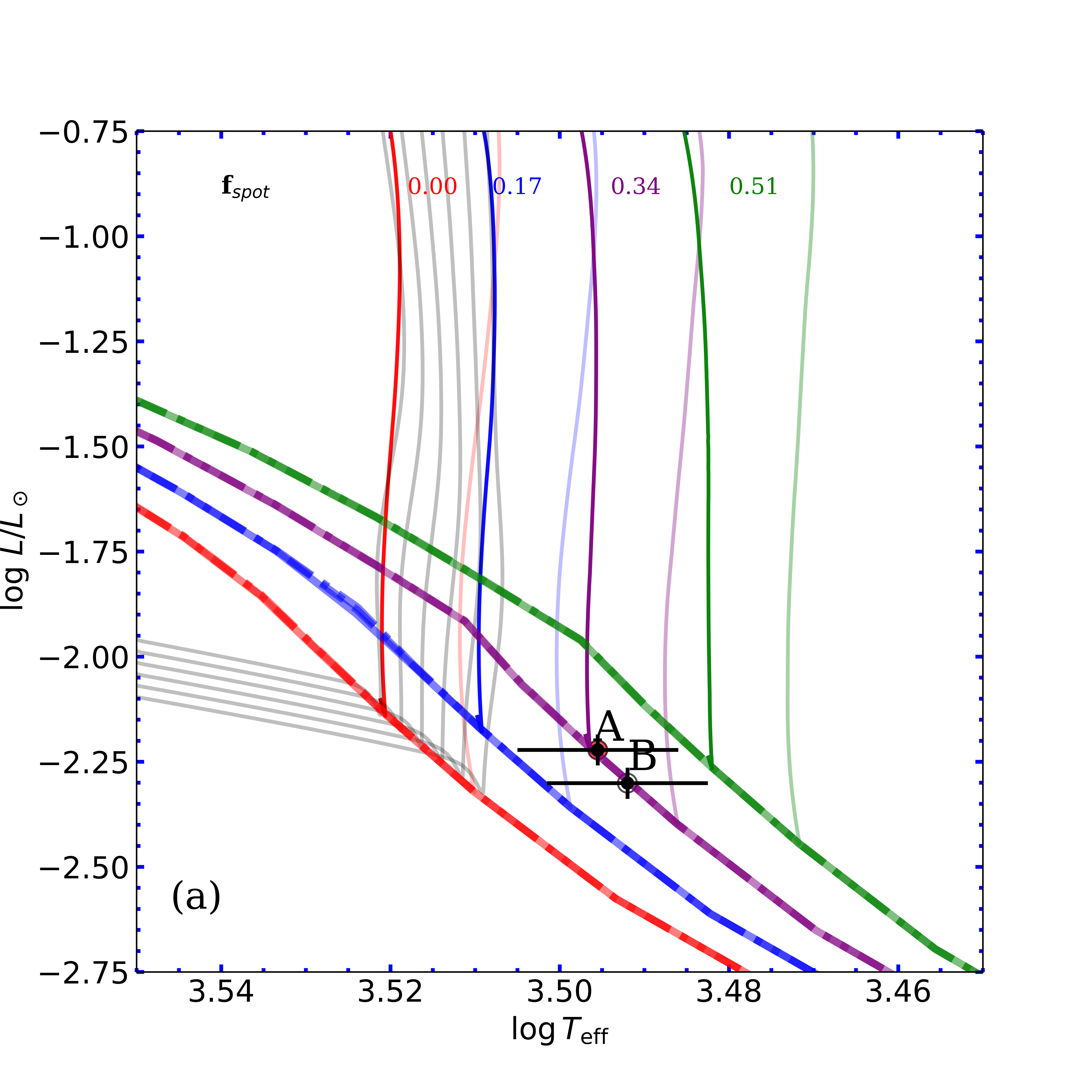}  
\includegraphics[width=0.45\linewidth]{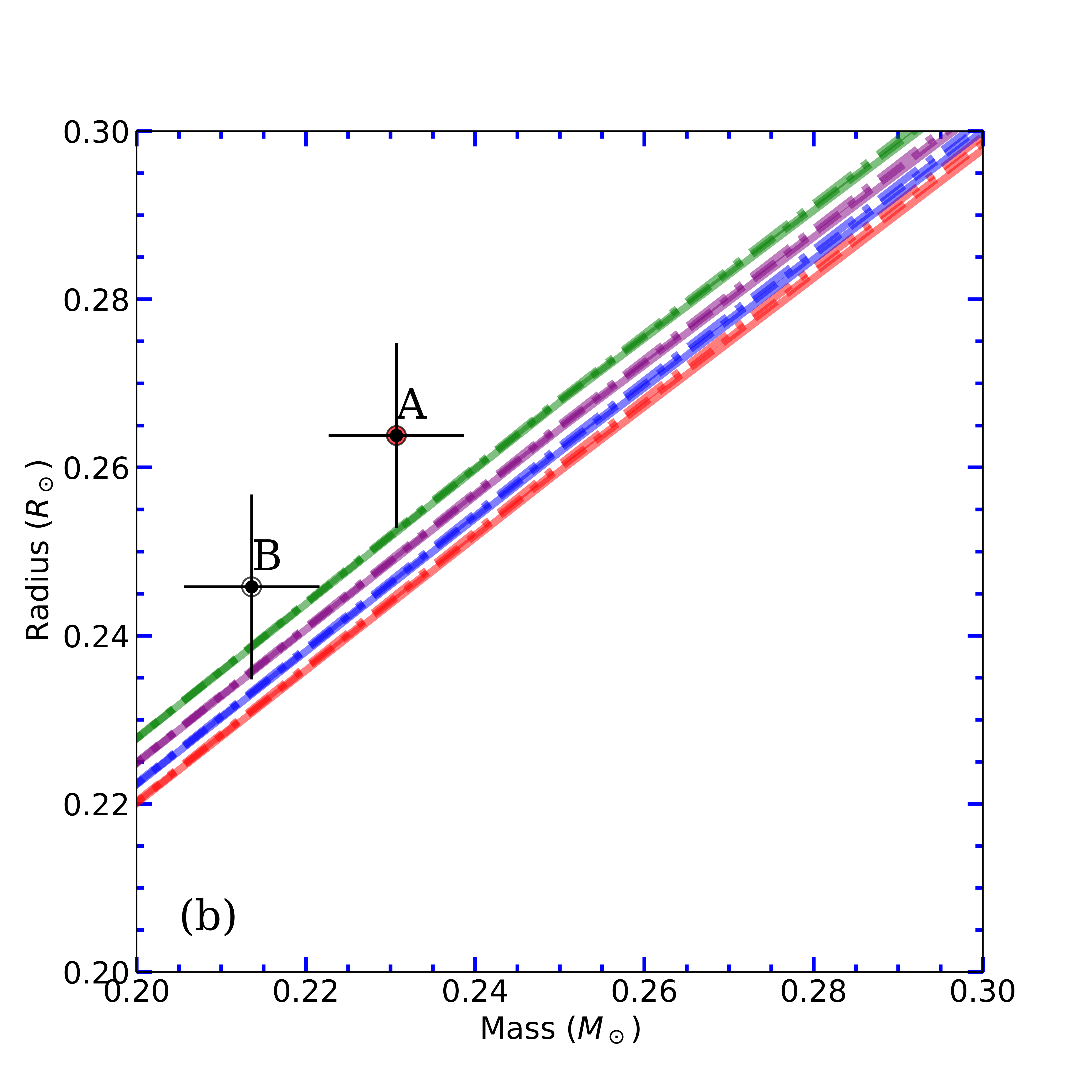} \\
 \vspace{-3mm}
\includegraphics[width=0.45\linewidth]{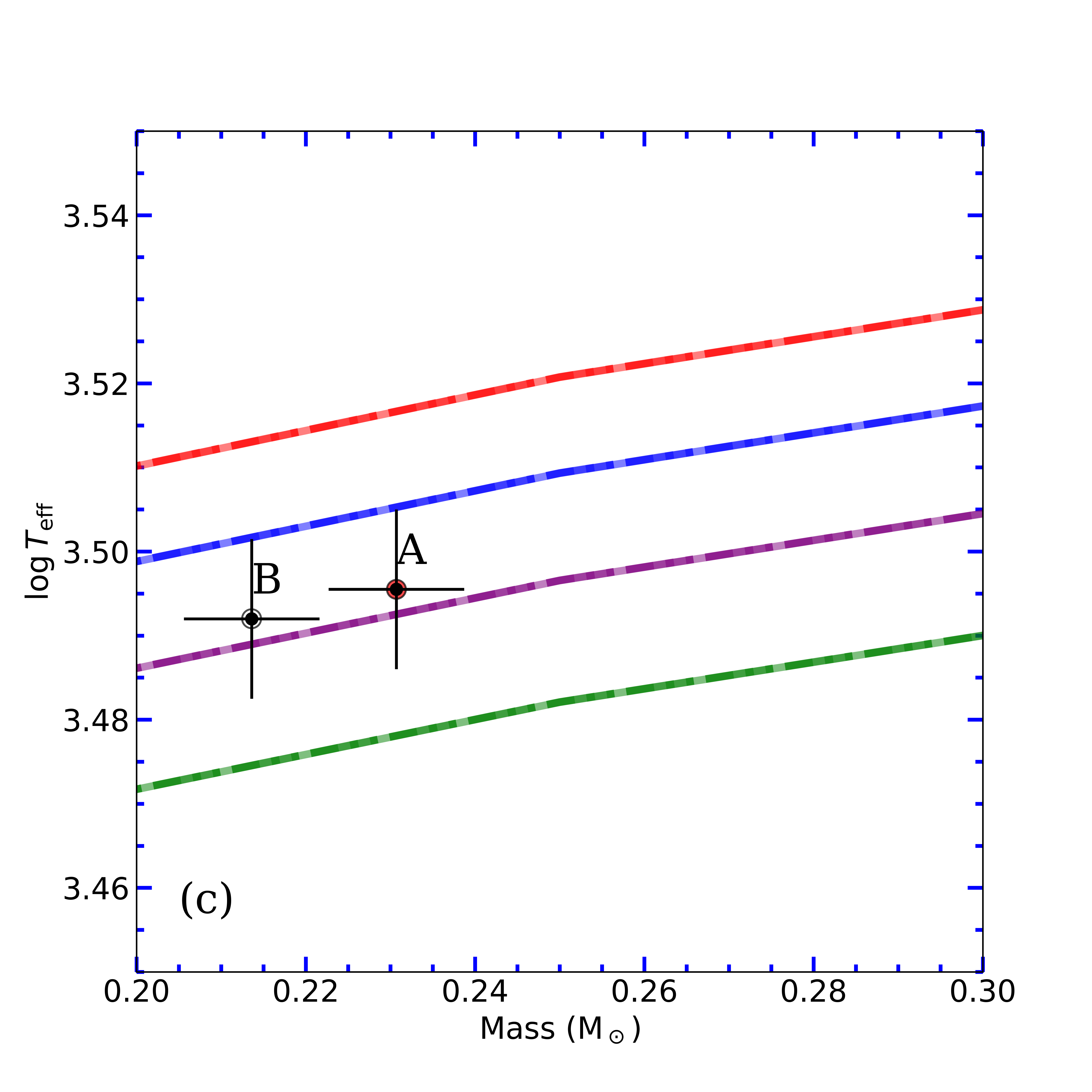} 
\includegraphics[width=0.45\linewidth]{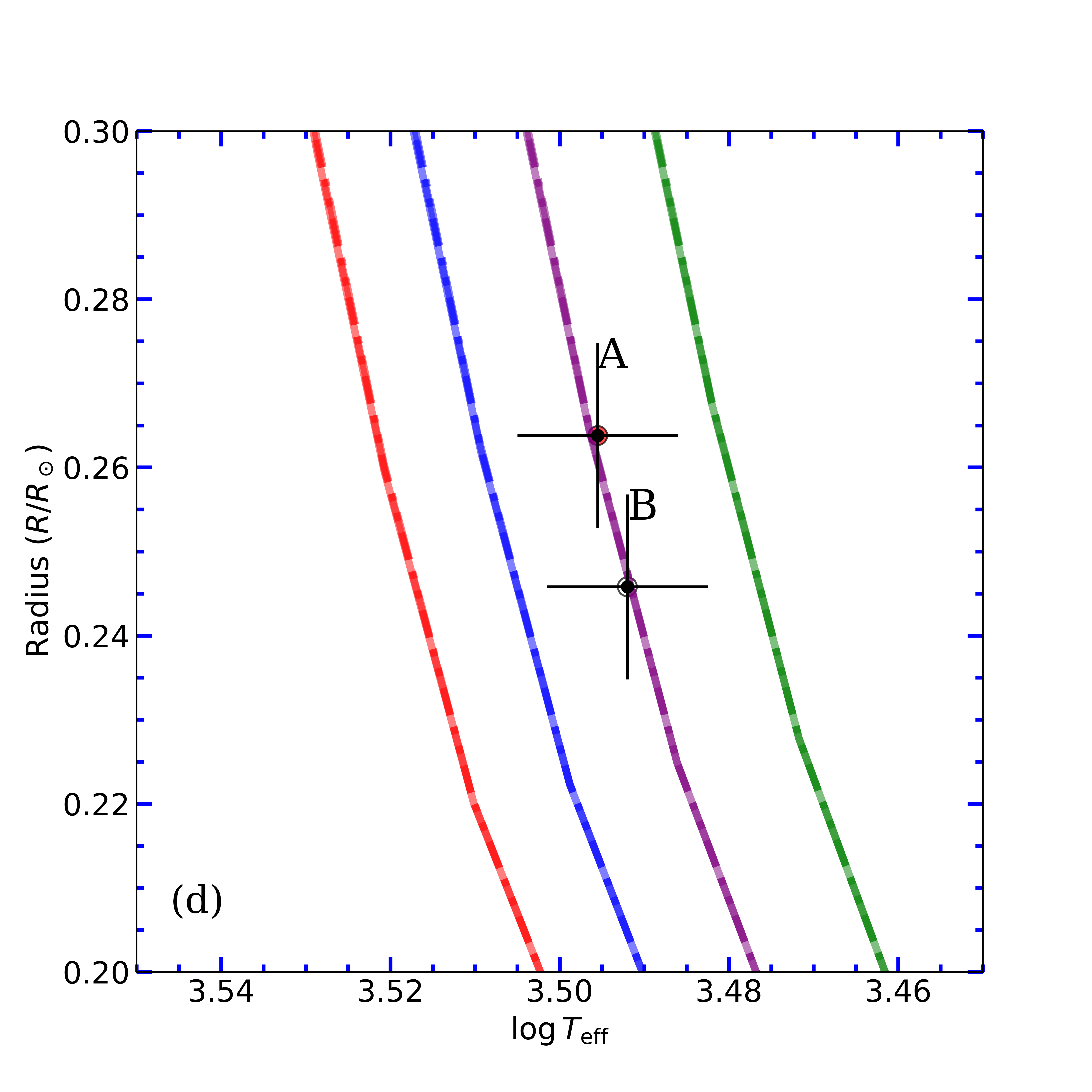} 
\caption{Evolutionary status for CM Dra in logT-logL (a), M-R (b), M-T (c) and T-R (d) planes. Our new results for CM Dra are shown as filled circles. SPOT models with f$_{\rm spot}$ values of 0.00 (red), 0.17 (blue), 0.34 (purple) and 0.51 (green) are considered in all panels. The isochrones are plotted with thick, bold lines in panel (a). In addition, \texttt{MIST} evolution tracks with $\rm{M_\odot}$ values of 0.20, 0.21, 0.22, 0.23, 0.24 and 0.25 are indicated with grey solid lines in the H-R diagram. As can be seen from the plots, models with mass loss are more successful in explaining the evolution of CM Dra.}
\label{fig:CMDra:HR}
\end{figure*}

We have shown a series of evolution models and isochrones taking different possible initial mass values of the system in the H-R, M-R, M-T$_{\rm eff}$ and T$_{\rm eff}$ - R planes in Figure~\ref{fig:CMDra:HR}. We plotted our newly obtained physical parameters for CM Dra, shown as filled circles in the figure. 
Among all the \texttt{SPOTS} models (f$_{\rm spot}$ =  0.00 (red), 0.17 (blue), 0.34 (purple) and 0.51 (green)) obtained, the models with f$_{\rm spot}$ parameter around 0.34 were found to be more consistent with the models. This result was also found to agree with the evolution models because of the medium-scale activity of CM Dra. In other words, when the effect of stellar activity is considered in the late-type stars, the results are more consistent with the observations. The isochrones are plotted with thick bold lines in Figure~\ref{fig:CMDra:HR}a. \texttt{MIST} evolution tracks (\citealt{choi16}; \citealt{dotter16}; \citealt{paxton15}) with $\rm{M_\odot}$ values of 0.20, 0.21, 0.22, 0.23, 0.24 and 0.25 are indicated with grey solid lines in the H-R diagram.  

Another situation that we should emphasize here is that the observational radius and temperature values observed in CM Dra-like systems (YY Gem, CU Cnc, etc.) differ from the expected theoretical values. This situation has been studied in detail by \cite{Rappaport2017MNRAS.467.2160R}, and the T$_{eff}$-M and R-M relations have been derived by comparing the existing observations with the theoretical models obtained. According to these relations, the expected component temperatures are 3310 K for the first star and 3280 K for the secondary component. As mentioned above, nonconservative stellar evolution models such as \cite{Yakut2005ApJ...629.1055Y}, \cite{Kalomeni2016ApJ...833...83K}, \cite{Eggleton2017MNRAS.468.3533E} and SPOT \citep{Somers2020ApJ...891...29S} approaches reasonably justify explaining the radius and temperature variations in such short-period active binary systems.

CM Dra is a particularly noteworthy astrophysical laboratory, distinguished by its unique combination of key characteristics. Its proximity to Earth facilitates remarkably precise measurements, while its status as a short-period, total eclipsing, and double-lined spectroscopic binary allows for highly accurate determinations of fundamental stellar parameters. The system's magnetic activity, manifesting as frequent flares and starspots, along with its apsidal-motion, offers insights into the dynamics of low-mass stars. The potential presence of a circumbinary planet, as evidenced by over half a century of extensive ground- and space-based observations, positions CM Dra as a preeminent candidate for the study of circumbinary planetary formation and evolution. The possible association with a distant white dwarf further enhances its significance, positioning CM Dra as a prime candidate for investigating multi-body dynamical interactions in a hierarchical stellar-planetary system. With high-precision measurements, long-term eclipse timing data, and a unique hierarchical structure, CM Dra represents a prime candidate for future astrometric and spectroscopic studies aimed at probing the long-term stability of planetary systems in multi-star environments.
Future high-resolution spectroscopic and astrometric observations, particularly with JWST and next-generation ground-based telescopes, will be essential in refining our understanding of this complex system, offering deeper insights into its dynamical architecture, stellar activity, and potential planetary companions. These observations will provide deeper insights into the evolutionary pathways of low-mass binary stars, the impact of stellar activity on planetary orbits, and the long-term dynamical stability of hierarchical planetary systems. Planetary systems with multiple stellar components have been documented in the literature, with notable examples including Kepler-64, a quadruple star system hosting a circumbinary planet \citep{Schwamb2013}, and Kepler-47, the first confirmed multi-planet circumbinary system \citep{Orosz2012}. Other confirmed cases, such as KOI-5 \citep{Wang2014}, KELT-4 \citep{Eastman2016}, and Gliese 667 \citep{Anglada-Escude2013}, demonstrate that planets can persist in dynamically complex environments where three or more stars are gravitationally bound. The CM Dra system, with its well-characterized eclipsing binary composed of two M-dwarfs and a widely separated white dwarf companion, represents a unique case. If the proposed circumbinary planet is confirmed, CM Dra would stand as a rare example of a planet-hosting hierarchical triple system, further underscoring the need for high-precision follow-up observations.

\section*{Acknowledgements}
The authors thank David Martin and Saul Rappaport for detailed reading of the paper and valuable comments and suggestions. This study was supported by the Scientific and Technological Research Council of T\"urkiye (T\"UB\.ITAK 112T766 and 117F188). In this study, observational data obtained within the scope of the project numbered 08ARTT150-35, conducted using the RTT150 Telescope at the TUG (T\"UB\.ITAK National Observatory, Antalya) site under the Türkiye National Observatories, have been utilized. The numerical calculations reported in this paper were partially performed at T\"UB\.ITAK ULAKB\.IM, High Performance and Grid Computing Center (TRUBA resources). KY thanks Churchill College for his fellowship.

\section*{Data Availability}

The data used in this study are given as tables online. In addition, TESS satellite data were used in some of the analyses and can be obtained from the MAST data archive at \href{https://mast.stsci.edu}{https://mast.stsci.edu}.



\bibliographystyle{mnras}
\bibliography{cmd} 



\bsp	
\label{lastpage}
\end{document}